\documentclass[fleqn,usenatbib]{mnras}

\usepackage{newtxtext,newtxmath}
\usepackage[T1]{fontenc}

\DeclareRobustCommand{\VAN}[3]{#2}
\let\VANthebibliography\thebibliography
\def\thebibliography{\DeclareRobustCommand{\VAN}[3]{##3}\VANthebibliography}

\usepackage{graphicx}	% Including figure files
\usepackage{amsmath}	% Advanced maths commands
\usepackage{todonotes}
\usepackage{float}
\usepackage{multirow}

\title[Comparing peaks between TNG and BCM]{Comparing weak lensing peak counts in baryonic correction models to hydrodynamical simulations}

\author[M. E. Lee \& T. Lu et al.]{
Max E. Lee,$^{1}$\thanks{E-mail: mel2260@columbia.edu}
Tianhuan Lu,$^{1}$
Zoltán Haiman,$^{1}$
Jia Liu,$^{2}$
Ken Osato$^{3,4}$
\\
$^{1}$Department of Astronomy, Columbia University, New York, NY 10027, USA\\
$^{2}$Kavli IPMU (WPI), UTIAS, The University of Tokyo, Kashiwa, Chiba 277-8583, Japan\\ 
$^{3}$Center for Gravitational Physics,
Yukawa Institute for Theoretical Physics, Kyoto University,\\
Kitashirakawa Oiwakecho, Sakyo-ku, Kyoto 606-8502, Japan\\
$^{4}$LPENS, D\'epartement de Physique, \'Ecole Normale Sup\'erieure,
Universit\'e PSL, CNRS, Sorbonne Universit\'e, Universit\'e de Paris,\\
24 rue Lhomond, 75005 Paris, France\\
}

\date{Accepted XXX. Received YYY; in original form ZZZ\\Report number:YITP-21-162}

\pubyear{2021}

\begin{document}
\label{firstpage}
\pagerange{\pageref{firstpage}--\pageref{lastpage}}
\maketitle
\defcitealias{Arico2020}{A20}

% Abstract of the paper
\begin{abstract}
Next-generation weak lensing (WL) surveys, such as by the Vera Rubin Observatory's LSST, the {\it Roman} Space Telescope, and the  \textit{Euclid} space mission, will supply vast amounts of data probing small, highly nonlinear scales. Extracting information from these scales requires higher-order statistics and the controlling of related systematics such as baryonic effects. To account for baryonic effects in cosmological analyses at reduced computational cost, semi-analytic baryonic correction models (BCMs) have been proposed. Here, we study the accuracy of BCMs for WL peak counts, a well studied, simple, and effective higher-order statistic. We compare WL peak counts generated from the full hydrodynamical simulation IllustrisTNG and a baryon-corrected version of the corresponding dark matter-only simulation IllustrisTNG-Dark. We apply galaxy shape noise expected at the depths reached by DES, KiDS, HSC, LSST, \textit{Roman}, and \textit{Euclid}. We find that peak counts in BCMs  are (i) accurate at the percent level for peaks with ${\rm S/N}<4$, (ii)  statistically indistinguishable from IllustrisTNG in most current and ongoing surveys, but (iii) insufficient for deep future surveys covering the largest solid angles, such as LSST and {\it Euclid}. We find that BCMs match individual peaks accurately, but underpredict the amplitude of the highest peaks. We conclude that existing BCMs are a viable substitute for full hydrodynamical simulations in cosmological parameter estimation from beyond-Gaussian statistics for ongoing and future surveys with modest solid angles.  For the largest surveys, BCMs need to be refined to provide a more accurate match, especially to the highest peaks.
\end{abstract}

% Select between one and six entries from the list of approved keywords.
% Don't make up new ones.
% \begin{keywords}
% keyword1 -- keyword2 -- keyword3
% \end{keywords}

%%%%%%%%%%%%%%%%%%%%%%%%%%%%%%%%%%%%%%%%%%%%%%%%%%

%%%%%%%%%%%%%%%%% BODY OF PAPER %%%%%%%%%%%%%%%%%%
\section{Introduction}
\label{sec:introduction}

Weak gravitational lensing (WL) measures the correlated distortions in the shapes of distant galaxies due to light deflections induced by gravitational potentials along the line of sight direction (see, e.g. \citealt{bartelmann01, Hoekstra08, kilbinger15} for reviews). Because of its sensitivity to the clustering of matter, WL is a promising cosmological probe. WL has already been shown to be a powerful tool for constraining parameters of the standard $\Lambda$CDM model, in particular the total matter density parameter, $\Omega_\mathrm{m}$, the amplitude of matter fluctuations $\sigma_8$, and their commonly used combination $S_8 = \sigma_8(\Omega_\mathrm{m}/0.3)^{0.5}$ \citep{Hikage19, Hamana20, DES21}.

The most common approaches to WL analysis rely on quantities related to the two-point correlation function of the shear field, or its Fourier transform, the power spectrum \citep{heymans13, kitching14, hildebrandt17, Hikage19}. However, it has been found that these statistics do not capture all the information from the non-linear and non-Gaussian regime of the WL field. This has led to an array of work attempting to extract this additional information, including non-Gaussian $N$-point statistics such as the 3-point function and its Fourier transform the angular bispectrum \citep{Fu14, coulton19, Halder21, Halder+2022},  Minkowski functionals  \citep{Maturi+2010,Munshi11, Kratochvil+2012, petri13, Marques2019, Gatti21}, convergence peak and minima statistics \citep{jain00, yang11, Liu15, kacprzak16, Li2019, Martinet2021, Zurcher2021, Harnois2021}, as well as machine learning methods \citep{Gupta18, Fluri19, ribli19, Matilla20, Lu21} and related techniques~\citep{Cheng+2020}.

Peak statistics have been shown to be a simple and effective non-Gaussian statistic of the WL field for constraining parameters and testing cosmological models. They represent pixels in the WL field that are larger in value when compared to their eight nearest neighbors \citep{jain00, Dietrich10, Kratochvil10,Liu15}. These peaks are correlated with massive individual dark matter haloes, or with numerous intermediate-mass haloes, making them sensitive to the structural evolution and history of the universe \citep{yang11,kilbinger15, Osato2015, liu16, Osato21}. It has recently been found in \cite{sabyr21} that beyond haloes, peaks are also sensitive to inter-halo matter. 

With future surveys covering $18,000\,\mathrm{deg}^2$ and galaxy number densities of $50\,\mathrm{arcmin}^{-2}$ \citep{LSST09, Euclid11}, systematic effects, such as intrinsic alignments of galaxy shapes \citep{schndeider10, sifon15}, and baryonic physics \citep{Jing06, Chisari19} will dominate the errors on small scales. As beyond-Gaussian statistics, such as peak counts, are particularly sensitive to small scales,  it is critical to understand such systematics for these statistics.

In this work, we focus our attention on the influence of baryonic physics on weak lensing peak statistics. Baryonic effects can be incorporated by hydrodynamical simulations that follow subgrid prescriptions calibrated to observations \citep{schaye10, McCarthy17, Volker10, Nelson2019}, by simpler semi-analytical baryonic correction models \citep{schndeider10, Schneider19, Arico2020}, or through machine learning methods \citep{camels20, Dai21, Lu2021}. While hydrodynamical simulations which have been appropriately calibrated are ideal testing grounds for baryonic effects and cosmological analyses, they require expensive simulation suites.

The semi-analytical baryonic correction models (BCMs) are attractive for cosmological analyses, as they correct computationally cheaper dark matter-only simulations without the need to explicitly simulate expensive baryonic processes. In \cite{schndeider10}, \citet{Schneider19}, and \citet{Arico2020}, BCMs with increasingly complex semi-analytical prescriptions for baryonic effects, such as star formation, active galactic nuclei (AGN) feedback, and gravitationally bound and unbound gas have been proposed and compared with full hydrodynamical simulations. In particular, \cite{Arico2020} have recently shown that their BCM is able to reproduce the power spectra of the 3D matter field to within $1\%$ accuracy in a broad range of state-of-the-art hydrodynamical simulations. 
 
To date, BCMs have been calibrated and tested almost exclusively on two-point statistics. The exception is the recent study by \cite{Arico+2021}, who examined a joint fit of their BCM to the 3D matter power spectrum and bispectrum, and found a $<3\%$ agreement with hydrodynamical simulations. In this work, we perform a similar comparison, focusing on WL peak counts---a statistic that contains information from all orders. We seek to answer the following questions: 

\begin{enumerate}
    \item How well do peak counts from BCM match hydrodynamical simulations?
    \item What is the physical cause of any apparent deviations?
    \item Can peak counts from BCMs be used as a replacement for hydrodynamical simulations in current and future survey analyses?
    \item Can the agreement between BCM and hydrodynamical peak counts be improved?
\end{enumerate}
  
This paper is organised as follows. In \S~\ref{sec:methods} we describe the simulations used, our weak lensing field computations, and the application of the \cite{Arico2020} baryonic correction model. We then present our comparisons between peak counts derived from hydrodynamical and BCM weak lensing maps in \S~\ref{sec:Results}. In \S~\ref{sec:discussion} we consider causes for the deviations between BCM and hydrodynamical peak counts, and explore the influence of galaxy shape noise on our results. This allows a quantification of how much the BCM will deviate from a hydrodynamical simulation if used in current and future analyses. We further propose and test a simple solution---excluding the highest peaks---to help mitigate these discrepancies. We summarise our main results and the implications of this work in \S~\ref{sec:conclude}.

\section{Methods}\label{sec:methods}

In this section we discuss the hydrodynamical and dark matter-only simulations used to generate WL maps. We then review the BCM of \cite{Arico2020} and its application to dark matter-only simulations. We end by describing the process of ray-tracing and the extraction of peak statistics. 

\subsection{$N$-body and Hydro Simulations}
\label{sec:simulations}

In this work we use the highest-resolution run of the magneto-hydrodynamical and dark matter-only cosmological simulations from IllustrisTNG-300, and IllustrisTNG-300-Dark~\citep[][hereafter TNG and TNG-Dark, respectively]{Pillepich17, springel18,Nelson2019,naiman18,Marinacci18}. TNG uses the adaptive moving-mesh code \texttt{AREPO} \citep{Volker10} on a $(205\,h^{-1}\mathrm{Mpc})^3$ simulation box with $2500^3$ dark matter particles and $2500^3$ initial gas cells. While both TNG and TNG-Dark contain the same initial random seed, the subgrid model of TNG incorporates baryonic physics, such as AGN feedback, star-formation and evolution, as well as radiative cooling \citep{Pillepich17}.

For TNG, the mass of a dark matter particle is $2.98\times10^7\, h^{-1}\, M_\odot$ and an initial gas cell mass is $1.1\times 10^7\,h^{-1}\,M_\odot$, while TNG-Dark has a dark matter particle mass of $4.72\times 10^7\,h^{-1}\, M_\odot$. Both simulations start at a redshift of $z=127$  and use $100$ snapshots to arrive at $z=0$, though in our work we focus our attention on snapshots between $0\leq z\leq 2.5$, covering the range of redshifts in WL surveys.

\subsection{Baryonic Correction Model}
\label{sec:BCM}
\begin{figure}
    \centering
    \includegraphics[width=\textwidth/2]{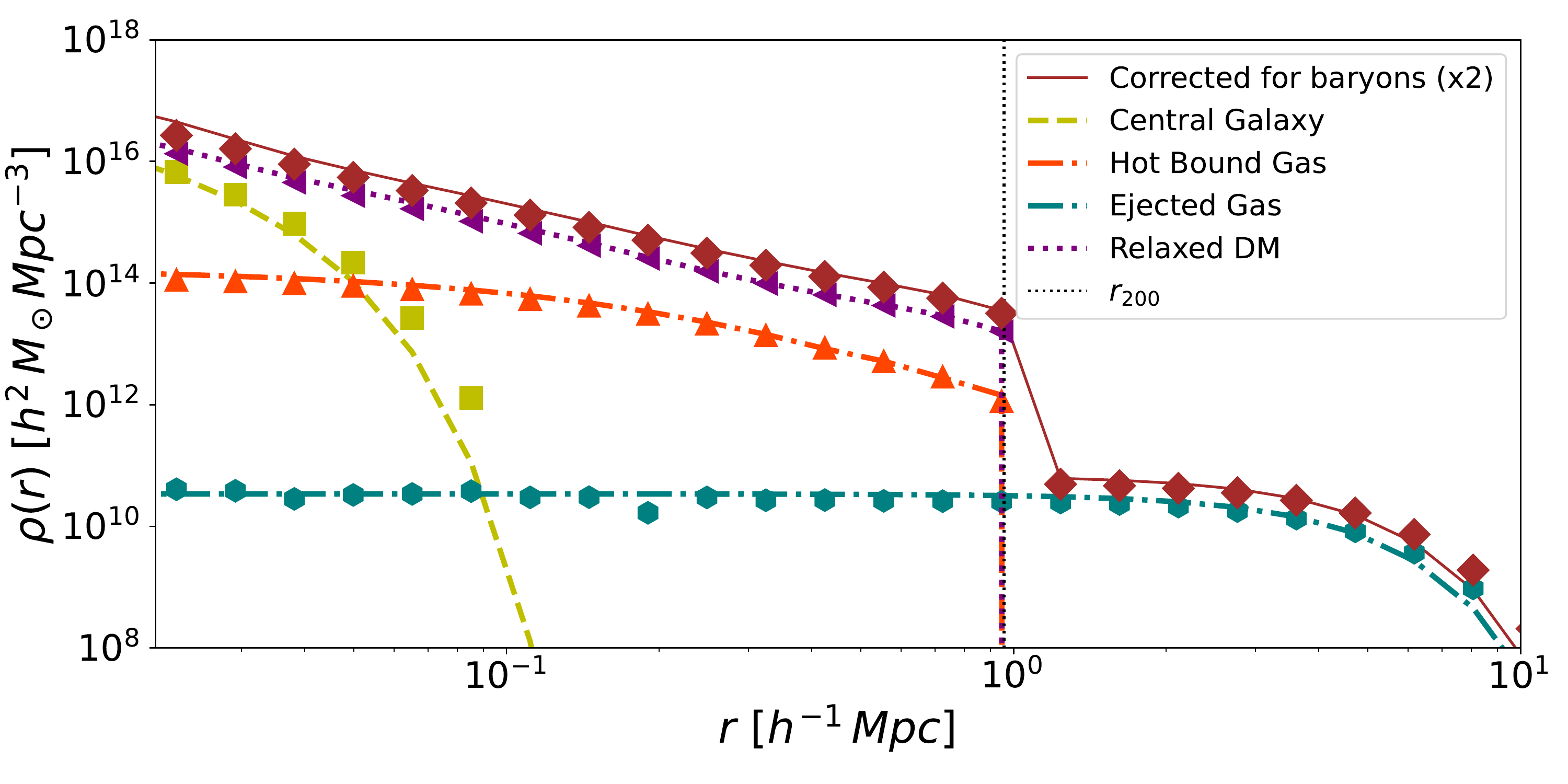}
    \caption{The density profiles of the four baryonic correction model (BCM) components and the total correction ($\times 2$ for visualization purposes) applied to a halo with mass $M_{200} \simeq 10^{14}\,h^{-1} M_\odot$. The lines here represent theoretical predictions for the BCM components, while the points show the measured values sampled from the halo after applying the BCM. This validates our implementation of the BCM outlined in \citetalias{Arico2020} (through the close match with Fig.~1 of \citetalias{Arico2020}).}
    \label{fig:density_prof}
\end{figure}
We use the four-parameter semi-analytical baryonic correction model outlined in \citet[][hereafter \citetalias{Arico2020}]{Arico2020}. \citetalias{Arico2020}'s corrections account for cooling of gas in halo centres, the ejection of gas from the haloes, hydrostatic equilibrium of hot bound gas, star formation, and adiabatic relaxation of dark matter due to baryons. Here we briefly summarise the key components and free parameters outlined in \citetalias{Arico2020} as well as in \cite{Lu2021}; interested readers are referred to these papers for more details.

In the BCM, density profiles of haloes are modified such that, 
\begin{equation}\label{eq:BCM_rho}
    \rho_\mathrm{DMO}(r) \to \rho_\mathrm{BCM}(r) = \rho_\mathrm{CG}(r) + \rho_\mathrm{HBG}(r) + \rho_\mathrm{EG}(r) +\rho_\mathrm{RDM}(r).
\end{equation}
We begin with a dark matter halo density profile $\rho_\mathrm{DMO}(r)$, which is then modified to be composed of a central galaxy component, $\rho_\mathrm{CG}(r)$, hot bound gas $\rho_\mathrm{HBG}(r)$, ejected gas, $\rho_\mathrm{EG}(r)$, and relaxed dark matter $\rho_\mathrm{RDM}(r)$, 
each with its own density profile. Once these density profiles are known, the enclosed-mass profile of the halo can be found by integrating Eq.~\eqref{eq:BCM_rho} to find $M_\mathrm{BCM}(r)$. We then invert this function to find $r_\mathrm{BCM}(M)$, and the corresponding correction of each simulated particle's radial position from the halo's centre of mass, 
\begin{equation}\label{eq:delta_r}
    \Delta r(M) = r_\mathrm{BCM}(M) - r(M).
\end{equation}

For each halo with a virial mass greater than $10^{12}\,h^{-1}\,M_\odot$ we extract all of the particles from its main sub-halo. This cut is reasonable because, as shown in \citetalias{Arico2020}, haloes with $M\leq 10^{13}\,h^{-1}\,M_\odot$ have an impact of $\leq 2\%$ on the total baryonic suppression at scales of $k\approx 5\,h\,\mathrm{Mpc}^{-1}$. For each sub-halo, we then fit the dark matter particles to a truncated NFW profile defined as \citep{NFW96} 
\begin{equation}
\begin{split}\label{eq:nfw}
\rho_\mathrm{NFW} = \begin{cases}
\frac{\rho_0}{(r/r_\mathrm{s})(1+r/r_\mathrm{s})^2}  \quad &{r\le r_{200}}\\
0  & {r>r_{200}}
\end{cases},
\end{split}
\end{equation}
where $\rho_0$ is the characteristic density of the halo, $r_\mathrm{s} = r_{200}/c$, $c$ is the concentration parameter, and $r_{200}$ is the radius at which the average density is $200$ times the critical density of the universe. We define $r_{200}$ as the halo radius and $M_{200}$ as the halo mass. The initial density profile $\rho_\mathrm{DMO}$ is then set to the NFW profile in Eq.~\eqref{eq:nfw}.

The central galaxy component follows a power-law with an exponential cutoff as in \cite{Kravstov18}. The slope and cutoff of the density profile is purely determined by the halo mass and radius, but the mass fraction $M_{\mathrm{galaxy}}/M_{\mathrm{halo}}$ is found from sub-halo abundance matching and uses the best-fit parameters found by \cite{Behroozi13}. To provide further flexibility, $M_1$, which is the mass of a halo with a galaxy mass fraction of $M_{\mathrm{galaxy}} / M_{\mathrm{halo}} = 0.023$ at $z=0$, is left as a free parameter.

We model hot bound gas in the halo as a fluid in hydrostatic equilibrium up to $\sqrt{5} r_{200}$ at which point it follows the slope of the truncated NFW profile of Eq.~\eqref{eq:nfw}. The fraction of bound gas is determined by the power law $(M_\mathrm{h}/M_\mathrm{c})^{\beta}$ where $M_\mathrm{h}$ is the halo mass, $M_\mathrm{c}$ is the mass at which half of the gas is retained inside of the halo, and $\beta$ is the slope of the gas fraction. Both $M_\mathrm{c}$ and $\beta$ are free parameters of the BCM.
 
Ejected gas assumes a Maxwell--Boltzmann velocity distribution with isotropic ejection. Particles are allowed to reach a maximum radius of $r_{\mathrm{ej}} = \eta \cdot 0.75r_\mathrm{esc}$ where $r_\mathrm{esc}$ is determined by considering the distance a gas particle with constant halo escape velocity, $v_{\mathrm{esc}}$, travels in half of the Hubble time \citep{schndeider10}, 
\begin{equation}
    r_{\mathrm{esc}} = 0.5t_{\mathrm{Hubble}} \sqrt{\dfrac{8}{3}G\Delta_{200}\rho_c}\,r_{200} \sim 0.5\sqrt{\Delta_{200}}\,r_{200}.
\end{equation}
In the BCM of \citetalias{Arico2020}, $\eta$ is the fourth and final free parameter. 

After considering the addition of baryonic effects, we introduce an adiabatic relaxation to the dark matter particles. This allows for positional adjustment given the gravitational interactions with the baryons. Thus regions of deep potentials, near the centre of haloes where we place the central galaxies, the dark matter particles contract. Regions near the edges of haloes, where ejected gas carries particles away, cause an expansion of the dark matter particles. 

We follow the original implementation of the BCM as in \citetalias{Arico2020}, in which we move each simulated DM particle in a given halo according to the BCM's semi-analytical prescription. Note that more efficient methods which rely on surface density projections \citep{Lu2021}, and emulators \citep{Arico20b} exist. For the purpose of this work, we find that the computational cost of individual particle displacement is not prohibitive, but for future work, we plan to implement these inexpensive simplifications.

We used $8$ nodes\footnote{Intel Xeon Phi 7250 KNL nodes on TACC's Stampede2, see \href{https://portal.tacc.utexas.edu/user-guides/stampede2}{https://portal.tacc.utexas.edu/user-guides/stampede2} for further node details} and $\sim1000$ core hours to correct each simulation snapshot from TNG-Dark on TACC's Stampede2. Using twenty TNG-Dark snapshots (see \S~\ref{sec:ray_tracing}), this corresponds to a total of $\sim20,000$ core hours. We have fully parallelised our BCM implementation using \textsc{mpi4py}\footnote{\url{https://github.com/mpi4py/mpi4py}}\citep{Dalcin21} which generates corrected maps far more efficiently. For simplicity, in this study we adopt the parameters already identified to best fit the 3D matter power spectrum in \citetalias{Arico2020}, as listed in Table~\ref{tab:bf_params}. In principle, optimal parameters could instead be found by fitting the WL peak counts directly. This could potentially improve the BCM's accuracy, but would require substantially more computational resources and work. We save this investigation for a future study, and simply note here that our results below are conservative.

As a test of our implementation, we compare the density profiles of each BCM component in Fig.~\ref{fig:density_prof}.  This plot gives a good intuition as to what the BCM is doing in general. The central galaxy component causes a sharp density increase at small radius which exponentially falls off, the warm gas then dominates the baryonic effects up to the halo edge at $r_{200}$. Near the outskirts of the halo, the ejected gas becomes dominant, causing particles originally inside the $r_{200}$ boundary to become expelled.  Our Fig.~\ref{fig:density_prof} is in excellent agreement with Figure~1 in \citetalias{Arico2020}, validating our BCM implementation in TNG-Dark.
\begin{table}
    \centering
    \begin{tabular}{|c|c|c|}
    \hline
    \hline
         Parameter & Value & Meaning\\
         \hline
         \hline
         $M_1 \, [10^{11} \mathrm{M}_\odot h^{-1}]$ & 0.22  & $M_{200}$ of halo with galaxy-mass fraction\\
          & & of $0.023$ at $z=0$\\
         $M_\mathrm{c} \, [10^{14} \mathrm{M}_\odot h^{-1}]$ & 0.23 & $M_{200}$ of halo with half gas bound\\
         $\beta$ & 4.09 & slope of gas fraction {\it vs.} halo mass\\
         $\eta$ & 0.14 & Maximum distance of ejected gas in units\\
         & & of $r_{\mathrm{esc}}$\\
         \hline
    \end{tabular}
    \caption{The parameter values used for the baryonic correction model. These are adopted from the fit in \citetalias{Arico2020} to the 3D matter power spectrum, which have been shown to yield $\leq 1\%$ errors.}
    \label{tab:bf_params}
\end{table}

\subsection{Ray-Tracing}
\label{sec:ray_tracing}

\begin{figure*}
    \centering
    \includegraphics[width=\textwidth]{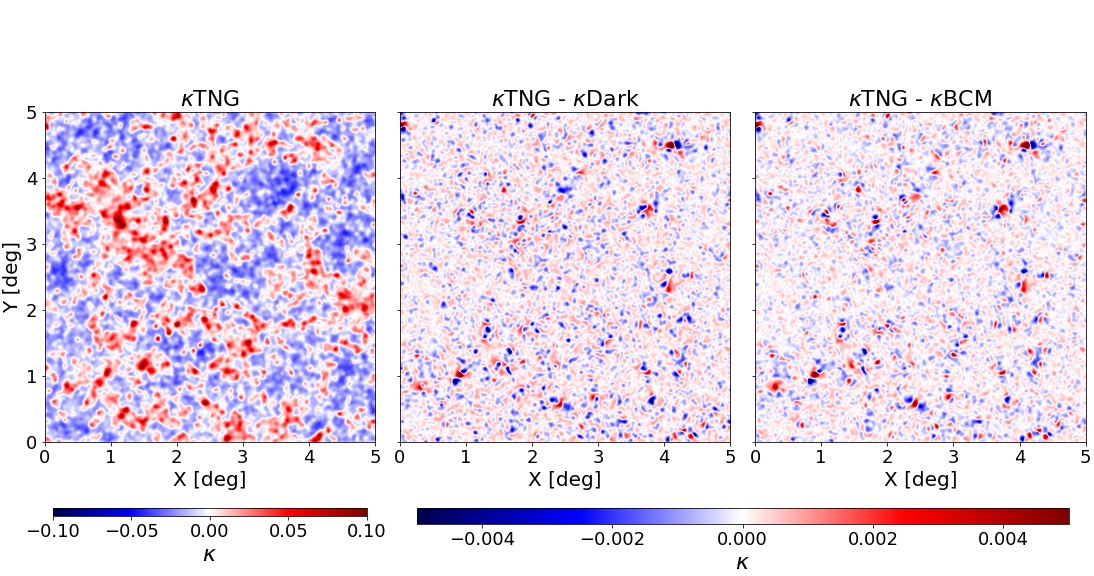}
    \caption{We show a typical $\kappa$TNG map at $z=1.5$ (\emph{left panel}), as well as the difference between a $\kappa$TNG map and $\kappa$Dark (\emph{middle panel}) and $\kappa$BCM (\emph{right panel)} that were generated with the same initial seeds. In the right two panels, we see a blue-red dipole pattern frequently occurring. We interpret this as being due to baryonic effects outside of haloes in TNG causing a shift of the centres of large haloes not captured by $\kappa$Dark and $\kappa$BCM (see \S~\ref{sec:discussion} for discussion).
    }
    \label{fig:kappa_diffs}
\end{figure*}

Here we briefly summarise the process outlined and used in \cite{Osato21}, which generated weak lensing convergence maps ($\kappa$ maps) from the TNG and TNG-Dark simulations. We use twenty TNG snapshots in intervals of approximately equally spaced comoving distance, each with a comoving volume of $(205\, \text{Mpc}\,h^{-1})^3$. This allows the generation of a light cone with a $5\times 5 \,\mathrm{deg}^2$ coverage\footnote{There is a tiling of snapshots used for farther redshifts to cover the full extent of the light cone. See \cite{Osato21} for details.}. We place source galaxies at four redshifts corresponding to $z=0.4, 0.7, 1.0, 1.5$ and model lensing convergence using the multiple lens plane algorithm of \cite{jain2000}. 

In this algorithm, each snapshot is split into two projections spanning a comoving distance of $102.5\,\mathrm{Mpc}\,h^{-1}$. The projection is then pixelated onto a $4096\times 4096$ mesh, on which the 2D lensing potential and deflection angles are computed. The change in lensed position ($\beta$), due to the deflection angle at each plane, with respect to its initial position ($\theta$) then represents the standard distortion matrix in weak lensing,
\begin{equation}
    \begin{split}
        A_{ij}(\Vec{\theta}, \chi)  &= \dfrac{\partial \beta_i(\Vec{\theta}, \chi)}{\partial \theta}\\
        &= \begin{pmatrix}
        1- \kappa - \gamma_1 & -\gamma_2 +\omega\\
        -\gamma_2 - \omega & 1-\kappa+\gamma_1
        \end{pmatrix}
    \end{split}
\end{equation}
where $\kappa$ is the weak lensing convergence, $\gamma_1$ and $\gamma_2$ are the shear components, and $\omega$ represents rotation. We employ the ray-tracing scheme of \cite{Hilbert09} to trace the light rays back from $z=0$ to the source redshift through each lensing potential along the light path. This method has been shown in many past studies to be memory efficient and accurate for WL simulations.

The resultant suite, dubbed $\kappa$TNG, contains $20,000$ total pseudo-independent convergence maps, each with $1024 \times 1024$ pixels, half of which are from ray-tracing the TNG and the other half from TNG-Dark simulations. We emphasise that these maps are pseudo-independent because each map was generated with the same twenty snapshots from TNG (TNG-Dark, BCM), though the particles in each snapshot were randomly translated, rotated by $0$, $90$, $180$, or $270$ degrees, and flipped along each of the three axes prior to the generation of lensing potential planes. In $\kappa$TNG this process was applied $100$ times, followed by a repetition of this process to each lens-plane another $100$ times to generate $10000$ pseudo-independent realizations. This process of translation and rotation to generate independent maps has been studied in \cite{petri16} and applied in \cite{Osato21} providing statistically independent power spectra and peak counts. 

Because of the computational cost of performing ray-tracing, we do not compute $10,000$ BCM $\kappa$ maps ($\kappa$BCM maps), and instead generate $2,000$ pseudo-independent convergence maps for the purpose of suppressing sample variance.  For each of $\kappa$BCM map, we find the concordant map (same random seeds) from $\kappa$TNG and $\kappa$Dark giving us a total of $6,000$ maps. Prior to analysis, all maps are smoothed with a Gaussian filter defined as
\begin{equation}\label{eq:window}
    W(\theta) = \frac{1}{\pi \theta_\mathrm{G}^2}\exp\left(-\frac{\theta^2}{\theta_\mathrm{G}^2}\right),
\end{equation}
where $\theta_\mathrm{G} = 1 \,\text{arcmin}$. This filter removes spurious small-scale noise, and was used in \cite{Osato21}. In the cases described in \S~\ref{sec:discussion}, we apply noise prior to smoothing.  We illustrate in Fig.~\ref{fig:kappa_diffs} a $\kappa$TNG map, and the differences $\kappa\mathrm{TNG} - \kappa\mathrm{Dark}$, and $\kappa\mathrm{TNG} - \kappa\mathrm{BCM}$ but save a discussion of this figure for \S~\ref{sec:discussion} below.

\subsection{Peak Statistics}
\label{sec:peak_stats}

We are interested in exploring the non-Gaussian statistic of peak counts from $\kappa$ maps. Peak counts are defined as the number of pixels with a given $\kappa$ value, that are greater than their surrounding eight neighbors. We explore peak counts from noiseless maps, and maps with Gaussian shape noise added to each pixel with a mean $\mu=0$ and variance 
\begin{equation}\label{eq:sigma}
    \sigma^2 = \frac{\sigma_e^2}{2n_\mathrm{gal} A_\mathrm{pix}}.
\end{equation}
Here $\sigma_e=0.4$ is the assumed mean intrinsic ellipticity of galaxies, $n_\mathrm{gal}$ is the surface number density of lensed galaxies, and $A_\mathrm{pix}$ is the solid angle of a pixel. We consider noise levels corresponding to current surveys with observed $n_\mathrm{gal} =10$ or $20 \,\mathrm{arcmin}^{-2}$, and future surveys with $n_\mathrm{gal}=30$ and  $50\,\mathrm{arcmin}^{-2}$.

We denote the average peak counts over all realizations ($N=2000$) of a method as $\langle\mathbf{n}\rangle$. Throughout this work, we represent our peak count histograms as a function of their significance in convergence maps, or ${\rm S/N}$.  Here the ``noise'' $N$ is taken as the root-mean-square $\kappa$ value ($\kappa_{\mathrm{rms}}$) of convergence maps, both noisy and noiseless, averaged over all $2000$ realizations. We show in Table~\ref{tab:rms_values} the $\kappa_{\mathrm{rms}}$ values from $\kappa$BCM and $\kappa$Dark in noiseless convergence maps for a range of redshifts. Choosing any of the columns in Table~\ref{tab:rms_values} is sufficient and only effects the $x$-axis scale by $<$ a percent level. This justifies the use of only $\kappa$TNG's $\kappa_{\rm rms}$ values throughout this work. $\kappa_{\rm rms}$ changes with noise level and redshift, and for each we compute the associated average $\kappa_{\rm rms}$ from $\kappa$TNG to find the S/N. We create 18 equally spaced bins from $-1.5 \leq {\rm S/N} \leq 7.5$ for each peak count histogram.

\begin{figure}
    \centering
    \includegraphics[width=\textwidth/2]{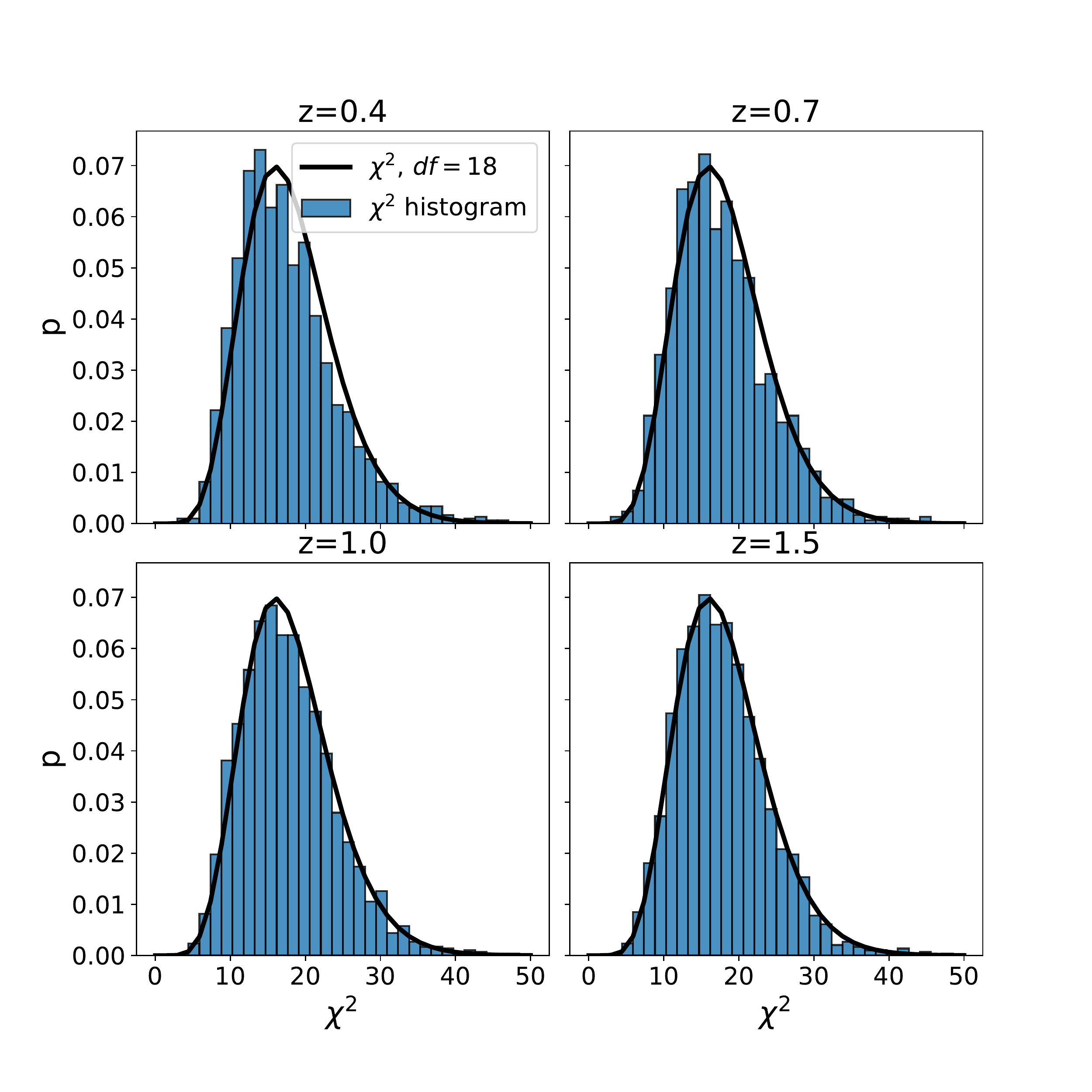}
    \caption{The distribution of the $\chi^2$ statistic defined in Eq.~\eqref{eq:chi2}, evaluated within the $\kappa$TNG simulation suite.  This statistic follows a $\chi^2$ distribution for 18 degrees of freedom (using 18 peak-count bins), as expected if the peak counts followed a multivariate Gaussian distribution. This allows us use $\chi^2$ to quantify the significance of the difference between the $\kappa$BCM and $\kappa$TNG peak-count histograms.}
    \label{fig:chi2-distribution}
\end{figure}

The significance of difference in peak counts between two methods, such as between the $\kappa$BCM and $\kappa$TNG, depends on two quantities: (i) the difference in the expectation value of the peak counts histogram $\Delta\langle\mathbf{n}\rangle$, and (ii) the peak counts covariances $\mathbf{C}$. The latter can be defined in either group of maps by

\begin{equation}\label{eq:cov}
    C_{ij} = \frac{1}{N-1} \sum_{k=1}^{N}
    \left(n_{i}^{(k)} - \langle n_i\rangle\right)\left(n_{j}^{(k)} - \langle n_j\rangle\right),
\end{equation}
where $N=2000$ denotes the number of realizations. For simplicity, we take the peak counts in the $\kappa$TNG maps to calculate the covariances here. We have found that evaluating the covariance matrix in the $\kappa$TNG or the $\kappa$BCM suites yields nearly identical $\chi^2$ values, and makes no difference to our results. Since the peak counts histogram generally follows a multivariate normal distribution~\citep{Gupta18}, the quantity
\begin{equation}\label{eq:chi2}
    \chi^2 \equiv {\left(\Delta\langle\mathbf{n}\rangle\right)}^T\, \hat{\mathbf{C}}^{-1}\,\Delta\langle\mathbf{n}\rangle,
\end{equation}
is a $\chi^2$ test (degrees of freedom $d=18$) for the difference between two peak counts histograms drawn from both methods, where
\begin{equation}
    \hat{\mathbf{C}}^{-1} = \frac{N-d-2}{N-1}\mathbf{C}^{-1},
\end{equation}
and prefactor $(N-d-2)/(N-1)$ is used to debias the precision matrix estimation \citep{hartlap07}. In Fig.~\ref{fig:chi2-distribution}, we validate that our definition of $\chi^2$ in Equation~\eqref{eq:chi2} indeed follows a $\chi^2$ distribution with $d=18$ when peak counts are sampled from $\kappa$TNG maps, i.e. $\Delta\langle\mathbf{n}\rangle$ is replaced by $\left(\mathbf{n}-\langle\mathbf{n}\rangle\right)$. This allows us to directly interpret the $\chi^2$ as the confidence at which one can reject the null hypothesis that the average value of the $\kappa$BCM peak histogram was drawn from the $\kappa$TNG distribution.  In particular, the standard 1, 2, and 3$\sigma$, or 68.27, 95.45 and 99.73\% confidence levels correspond to probabilities of finding $\chi^2\leq 20.28$, 29.24 and 39.17, respectively, by random chance, for a $\chi^2$ distribution with 18 degrees of freedom. 

The above calculation only applies to comparing individual peak counts histograms of $5\times5\,\mathrm{deg}^2$ maps from different methods, which corresponds to a $25\,\mathrm{deg}^2$ survey. To derive the significance of difference for larger surveys, we shall generalize the calculation. We note that the peak counts histogram of a survey with a area $A$ is equal to the average of $A/(25\,\mathrm{deg}^2)$ draws of $25\,\mathrm{deg}^2$ histograms, which does not change the expectation value $\langle\mathbf{n}\rangle$ but scales the covariance matrix by
\begin{equation} \label{eq:cov_area}
    \mathbf{C}(A) = \left(\frac{A}{25\,\text{deg}^2}\right)^{-1} \mathbf{C}(25\,\mathrm{deg}^2),
\end{equation}
where $\mathbf{C}(25\,\mathrm{deg}^2)$ follows Equation~\eqref{eq:cov}. $\chi^2$ of this larger survey is modified accordingly:
\begin{equation} \label{eq:chi2_area}
    \chi^2(A) = \frac{A}{25\,\text{deg}^2}\,\chi^2(25\,\mathrm{deg}^2),
\end{equation}
where $\chi^2(25\,\mathrm{deg}^2)$ follows Equation~\eqref{eq:chi2}. Intuitively, when we switch from a small survey to a larger one, the peak counts will be of similar values but the uncertainties of the counts will become smaller; therefore, with the same absolute difference between $\kappa$BCM and $\kappa$TNG, our ability of distinguishing one method from the other is improved ($\chi^2$ is proportional to $A$). Hereafter, we refer to $\chi^2(A)$ as simply $\chi^2$ when it is calculated for each specific survey.

\subsection{Summary}

We summarise our process below. 
\begin{enumerate}
    \item Extract primary sub-halo from all haloes  with $M_{200}\geq 10^{12}\,h^{-1}\mathrm{M}_\odot$ from TNG-Dark, for $20$ snapshots between $z=0$ and $z\approx2.5$, using halo catalogs released with TNG-Dark
    \newline
    \item Fit NFW profiles (Eq.~\eqref{eq:nfw}) to each of the haloes to find $\rho_\mathrm{DMO}(r)$
    \newline
    \item Compute BCM density profiles $\rho_\mathrm{BG},\rho_\mathrm{EG},\rho_\mathrm{CG},\rho_\mathrm{RDM}$
    \newline
    \item Find the new density profile (Eq.~\eqref{eq:BCM_rho}) and apply radial shifts for each particle (Eq.~\eqref{eq:delta_r})
    \newline
    \item Compute 2000 $\kappa$ maps following the prescription of \cite{Osato21} using BCM corrected 3D matter fields
    \newline
    \item Add galaxy shape noise following Eq.~\eqref{eq:sigma}
    \newline
    \item Smooth all maps with a $\theta_\mathrm{G}=1\, \mathrm{arcmin}$ Gaussian filter  (Eq.~\eqref{eq:window})
    \newline
    \item Compute the peak counts as described in \S~\ref{sec:peak_stats} for $\kappa$TNG, $\kappa$Dark, and $\kappa$BCM
    \newline
    \item Calculate covariance matrices (Eqs.~\eqref{eq:cov} and \eqref{eq:cov_area}) and finally the $\chi^2$ (Eqs.~\eqref{eq:chi2} and \eqref{eq:chi2_area}) between the $\kappa$BCM and $\kappa$TNG peak count histograms.
\end{enumerate}

\section{Results}
\label{sec:Results}
\begin{figure}
    \centering
    \includegraphics[width=\textwidth/2]{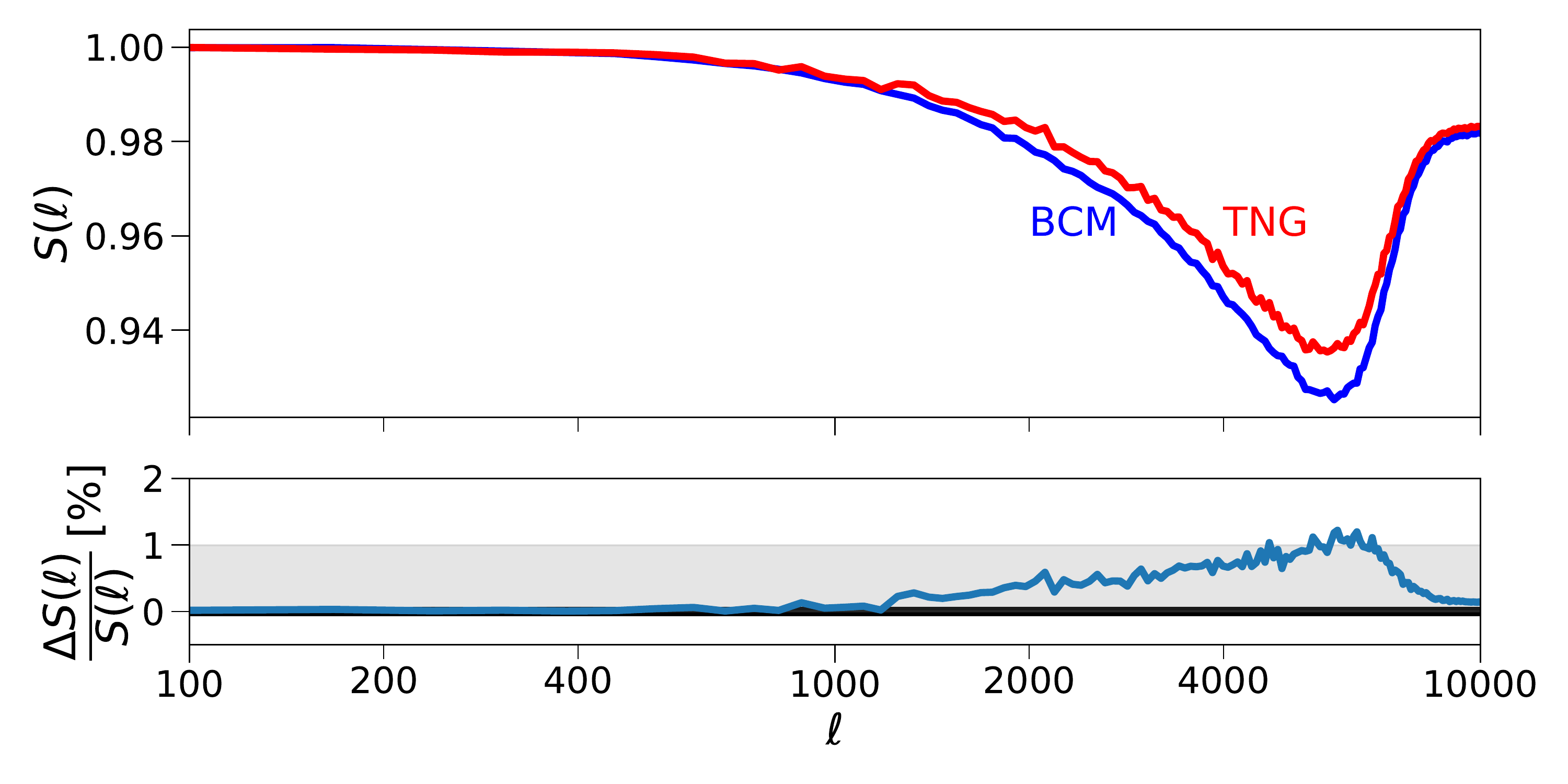}
    \caption{The suppression in the weak lensing convergence power spectrum defined as $S(\ell) = \langle P_{\mathrm{Baryons}}(\ell) \rangle/ \langle P_{\mathrm{Dark}}(\ell)\rangle $. \emph{Top panel:} suppression in both the full hydrodynamical simulation ($\kappa$TNG, in blue) and in our BCM implementation (in orange), finding good agreement between the two. \emph{Bottom panel:} the fractional difference between the two remains mostly below $\approx 1\%$. Both lines show averages over 2000 realizations, without noise added to the convergence maps.}
    \label{fig:kappa_power}
\end{figure}
\begin{figure*}
    \centering
    \includegraphics[width=\textwidth]{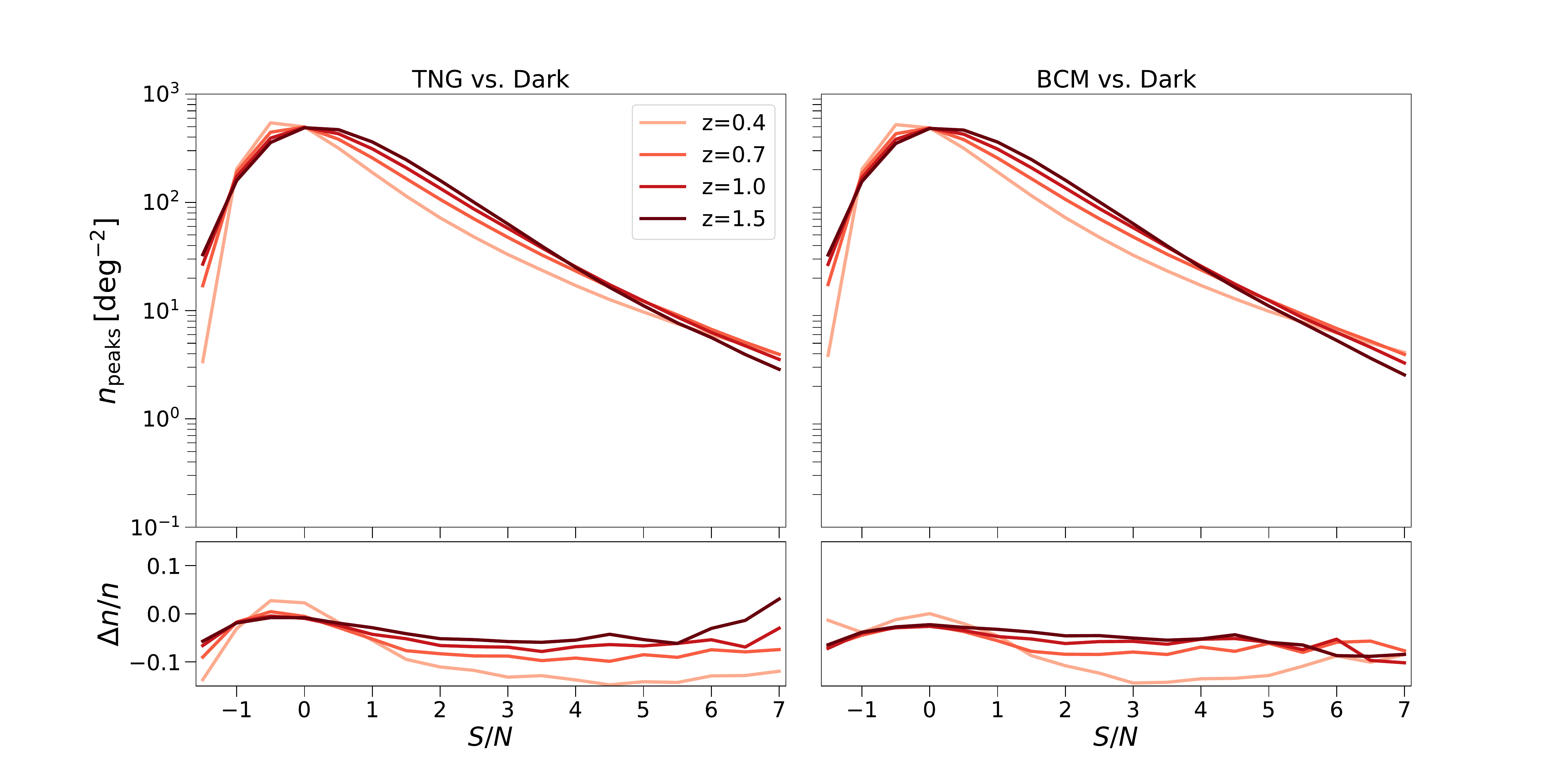}
    \caption{Peak histograms from noiseless $\kappa$TNG (\emph{upper left panel}) and $\kappa$BCM 
    (\emph{upper right panel}). In both cases, we show the deviations in peak counts compared to $\kappa$Dark (\emph{lower panels}). At ${\rm S/N}\gtrsim4$ we find that the BCM underpredicts the number of peaks when compared to $\kappa$Dark maps, whereas TNG leads to a higher number of peaks.
    }
    \label{fig:peak_hist}
\end{figure*}
\begin{table*}
    \centering
    \begin{tabular}{c|c|c|c|c|c|c|}
    \hline
    \hline
         Survey & Approx. $n_{\mathrm{gal}}\,[\mathrm{arcmin}^{-2}]$ & Approx. Area [$\mathrm{deg}^2$] & Approx. Median z\\ 
         \hline
         \hline
        {\it Roman}   &  $50$  &  $2200$ & $1.5 $\\
        LSST    &  $30$  &  $18000$ & $1.0$\\
        {\it Euclid}  &  $30$  &  $15000$ & $1.0$ \\
        HSC     &  $20$  &  $1500 $  & $0.7$\\
        DES     &  $10$  &  $5000 $  & $0.4$\\
        KiDS    &  $10$  &  $1350 $  & $0.4$\\
        \hline
    \end{tabular}
    \caption{A summary of the approximate survey depth, area, and median redshift we adopted to correspond to the six weak lensing surveys {\it Roman}, LSST, {\it Euclid}, HSC, DES and KiDS.}
    \label{tab:survey_details}
\end{table*}
In this section, we present results from weak lensing statistics compared between $\kappa$TNG, $\kappa$Dark, and $\kappa$BCM.  We first validate our BCM, ray-tracing, and peak count implementations through comparisons to previous work.  We then discuss the peak count histograms in $\kappa$TNG and in $\kappa$BCM.

As a validation step, we compute power spectra of each of the WL convergence maps, and find the average power spectrum suppression, defined as
\begin{equation}
    S(\ell) \equiv \frac{P_\mathrm{Baryons}(\ell)}{P_\mathrm{Dark}(\ell)},
\end{equation}
for both Baryons $=$ TNG and Baryons $=$ BCM. Here $P(\ell)$ indicates the average power spectrum over the 2000 realizations, measured using the package \textsc{nbodykit}\footnote{\url{https://github.com/bccp/nbodykit}}\citep{Hand18}. The results in Fig.~\ref{fig:kappa_power} show that the average suppressions in the $\kappa$BCM and $\kappa$TNG suites are very similar, with the absolute power spectra agreeing to within  $\sim1\%$ at all $\ell$. On the other hand, the $\kappa$BCM power spectrum is systematically slightly lower than $\kappa$TNG, which is likely caused by a somewhat excessive AGN feedback component in the BCM. This effect has been shown by \citetalias{Arico2020} and in \cite{Lu21} to dominate the power spectrum suppression on small scales when using BCM's. While in this work we use the BCM parameters fit to the 3D matter power spectrum from \citetalias{Arico2020}, in the future we will investigate directly fitting BCM parameters to other WL statistics. Here, we find that the $\kappa$BCM's difference from $\kappa$TNG is small and consistent with previous studies, validating our implementation \citep{Arico20b, Lu2021}.

Next, for each smoothed convergence map, we identify the peaks and measure the $\kappa_{\mathrm{rms}}$ values following \S~\ref{sec:peak_stats}. In Fig.~\ref{fig:peak_hist}, we show the peak count histogram obtained for $\kappa$TNG and $\kappa$BCM in the top panels, as well as their comparison to $\kappa$Dark in the bottom panels. We plot these histograms as functions of their significance (${\rm S/N}$) in $\kappa$ maps, where ${\rm S/N}\equiv\kappa_{\rm peak}/\kappa_{\mathrm{rms}}$. Throughout this work, we use approximate median redshifts, galaxy densities, and areas found by inspecting details of \cite{LSST09, DES21, HSC_2018, Euclid11, KiDs13, Roman15} (see Tab.~\ref{tab:survey_details}). Fig.~\ref{fig:peak_hist} serves two purposes: (i) we are able to match our results to \cite{Osato21}, validating our convergence map and peak count pipeline, and (ii) we can compare the deviations between peaks (\emph{lower panels}). We see qualitatively that the differences between the $\kappa$TNG {\it vs.} $\kappa$Dark histograms are well matched by $\kappa$BCM {\it vs.} $\kappa$Dark up to ${\rm S/N} \sim 4$. Above this value,  $\kappa$TNG has more peaks than $\kappa$Dark, yielding a positive slope, whereas the BCM systematically underpredicts the number of peaks, causing a negative slope. We explore the implications of this more in \S~\ref{sec:peak_locs_and_heights}

\begin{figure*}
    \centering
    \includegraphics[width=1\textwidth]{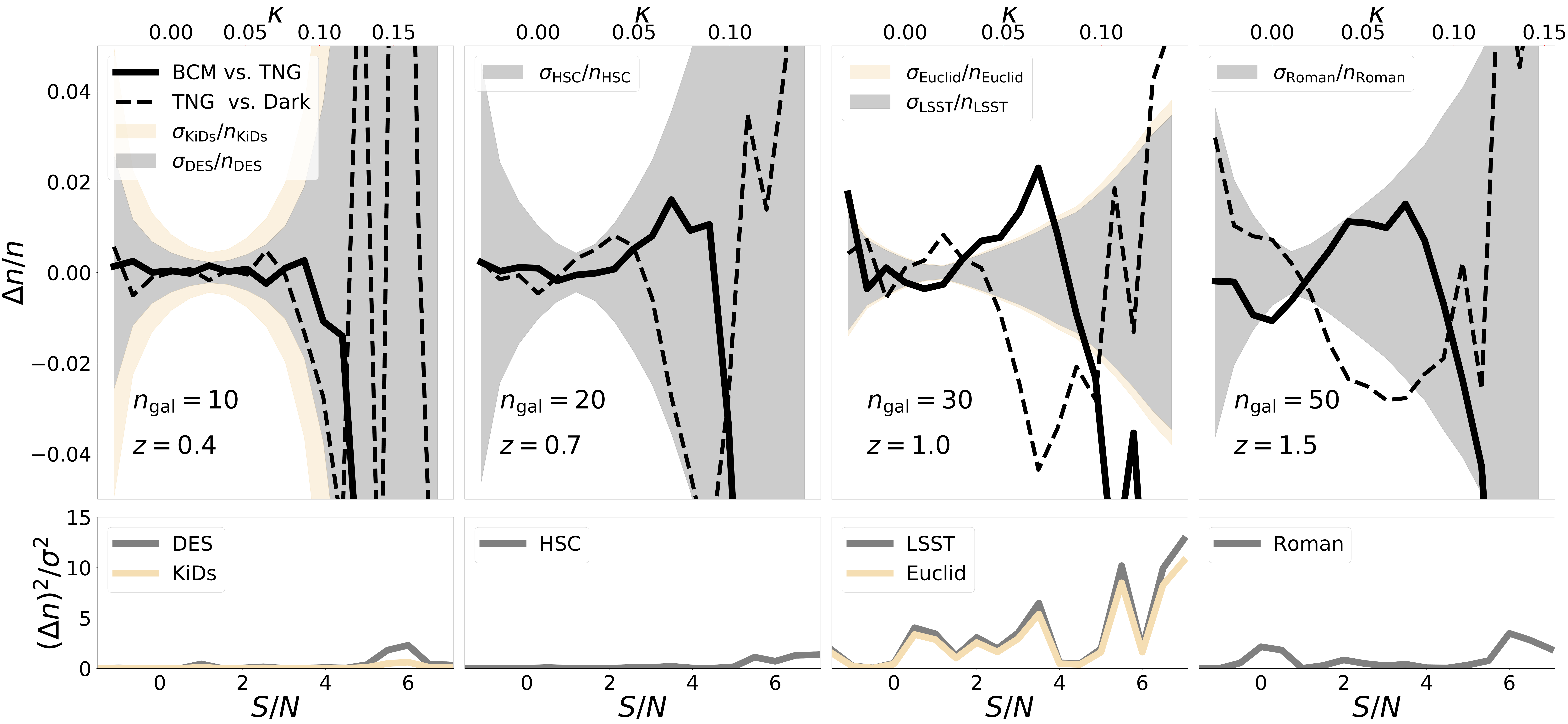}
    \caption{We compare the peak counts in $\kappa$BCM and $\kappa$TNG as a function of peak height significance (S/N; bottom labels) and $\kappa$ (top labels). The four panels show different combinations of median source galaxy redshift and noise level, roughly corresponding to the redshifts and galaxy number densities in different WL surveys, as labeled.  We find that the agreement between peaks from $\kappa$BCM and $\kappa$TNG is $<2\%$ up to ${\rm S/N}\approx 4$, beyond which the discrepancies increase sharply.    The shaded regions correspond to $1\sigma$ Poisson uncertainties in each bin for DES, HSC, {\it Euclid}, LSST, and {\it Roman}-like surveys (with the assumed areas listed in Table~\ref{tab:survey_details}), illustrating the significance of the deviations in peak counts. The bottom panels show the contribution to the $\chi^2$ from each bin (see text for a numerical calculation, taking into account correlations between bins). For reference, we also show the (much larger) differences between $\kappa$TNG and $\kappa$Dark (dashed curves).}
    \label{fig:BCM_TNG}
\end{figure*}

Next, we quantify how significant the deviations are between $\kappa$TNG and $\kappa$BCM peaks. We consider surveys with depths ranging from $10\,\mathrm{arcmin}^{-2} \leq n_{\mathrm{gal}} \leq 50 \, \mathrm{arcmin}^{-2}$ and with median redshifts ranging from $0.4\leq z\leq 1.5$.  We simplify our analysis by placing all galaxies at a single redshift and add shape noise following Equation~\eqref{eq:sigma}. We smooth and compute the average $\kappa_{\rm rms}$ and peak count histogram over all $2000$ map realizations. To visually illustrate the significance of the difference between the two histograms, we consider the Poisson uncertainty in each bin, scaled to the appropriate survey size. In Fig.~\ref{fig:BCM_TNG}, we then show the relative errors along with the Poisson uncertainties,  for depths, redshifts and solid angles roughly corresponding to those in six different WL surveys. The parameters we adopted for these surveys are listed in Table~\ref{tab:survey_details}. 

Fig.~\ref{fig:BCM_TNG} shows a $\sim2\%$ agreement between $\kappa$TNG and $\kappa$BCM peak statistics over the range $-1.5<{\rm S/N}<4$ for most galaxy densities considered. The shaded regions in this figure mark the Poisson uncertainties scaled to the rough areas of the DES, KiDs, HSC, {\it Euclid}, LSST, and {\it Roman} surveys. We further plot the relative difference between $\kappa$TNG and $\kappa$Dark, to show that the overall effect of the baryons is much larger than the errors introduced by using the BCM. It is clear from this visualization that $\kappa$BCM at lower redshifts and higher noise levels achieves greater relative accuracy than at high redshift and with low noise. Comparing the curves to the shaded areas also gives a rough estimate of the significance of the deviations in each S/N bin. We further include the corresponding $\kappa$ values which confirms a weak dependence of $\kappa_{\rm rms}$ with respect to $n_{\rm gal}$ when compared to the the dependence of $n_{\rm peaks}$ on $n_{\rm gal}$.

In the next section, we account quantitatively for  correlations between bins by computing the covariance matrix, the $\chi^2$ values, and the corresponding probabilities. We also study how  the significance of the $\kappa$BCM {\it vs.} $\kappa$TNG difference is impacted by the presence of shape noise.

\begin{table}
    \centering
   \begin{tabular}{lrrr}
   \hline
   \hline
        $z$ &       $\kappa_{\mathrm{rms}}^{\mathrm{TNG}}$ &        $\kappa_{\mathrm{rms}}^{\mathrm{BCM}}$ &       $\kappa_{\mathrm{rms}}^{\mathrm{Dark}}$ \\
        \hline
        \hline
        0.4 &  0.0040 &  0.0039 &  0.0041 \\
        0.7&  0.0069 &  0.0068 &  0.0070 \\
        1.0 &  0.0099 &  0.0099 &  0.0100 \\
        1.5 &  0.0136 &  0.0136 &  0.0137 \\
        \hline
        \end{tabular}
    \caption{The $\kappa_{\mathrm{rms}}$ values at the four different redshifts with no noise used in this work. We show $\kappa_{\mathrm{rms}}$ for $\kappa$TNG, $\kappa$BCM, and $\kappa$Dark to show how close they are, and to justify the use of $\kappa_{\mathrm{rms}}^{\mathrm{TNG}}$ as the ``noise'' $N$ in determining the significance (S/N) of peaks in $\kappa$ maps. Each value was computed by averaging the root-mean-square over all $2000$ convergence maps.}
    \label{tab:rms_values}
\end{table}

\section{Discussion}\label{sec:discussion}

In the previous sections we described the process of modifying the TNG simulations to account for baryons using the BCM of \citetalias{Arico2020}, how we ray-traced through this simulation, and the statistics measured from the resulting maps. The main result we found is that peak counts using the BCM agree at the percent level with the full hydrodynamical TNG simulation up to a peak height of ${\rm S/N} \approx 4$. In this section, we further quantify this result and discuss its implications. We also look more closely at matches between individual peaks at the map level, and assess whether the BCM's accuracy is sufficient for current and upcoming surveys.

\subsection{Accuracy of BCM as a function of survey depth and area}

The goal of using BCMs for WL peak count analysis is to avoid the large computational costs associated with high-resolution hydrodynamic simulations with baryonic physics. A natural question then is to what extent BCMs will be a viable alternative in the analyses of current and future surveys?  One could either attempt to simultaneously fit cosmological and baryonic parameters, or simply marginalize over the baryonic physics through the BCM parameters.

Fig.~\ref{fig:BCM_TNG} partly addresses this question. It shows that the systematic deviation in peak counts introduced by the BCM  stays within $2\%$ up to peak heights of ${\rm S/N}\approx 4$. Furthermore, this is within the Poisson uncertainty in each bin, except for the largest (nearly all-sky) surveys.  The shaded region in this plot represents the square root of the covariance matrix's diagonal elements. The ratio of the $\mathbf{\Delta n}$ curves to the width of the shaded area is therefore the contribution to the $\chi^2$ from each bin (corresponding to the diagonal elements in the full matrix product Eq.~\eqref{eq:chi2}). These are shown in the bottom panels, and reveal that the largest contributions to  $\chi^2$ are from peaks with ${\rm S/N}\gtrsim 4$ for all redshifts and noise levels. However, we note that for LSST and {\it Euclid}, at $z=1$ and $n_{\mathrm{gal}} = 30$, significant contributions to $\chi^2$ come from smaller peaks ($0\lesssim{\rm S/N}\lesssim 4$) as well.

We next include correlations between bins, i.e. compute the full $\chi^2$ including the off-diagonal terms of the covariance matrix in Eq.~\eqref{eq:chi2}. We compute peak histograms at each of the four redshifts $z=0.4, 0.7, 1$ and $1.5$ used in this work, roughly corresponding to the median redshift of current and future surveys. At each redshift, we add shape noise corresponding to $n_{\mathrm{gal}} = 10, 20, 30, 50 \,\mathrm{arcmin}^{-2}$, respectively, which roughly represent the depths of these surveys (see Table~\ref{tab:survey_details}). All galaxies are placed at a single source plane at the median redshift for simplicity. We adopt the same binning as in \S~\ref{sec:peak_stats} and follow the procedure outlined in that section to compute $\chi^2$ for each combination of noise level, source redshift, and survey area.
Note that $\chi^2$ is simply proportional to survey size.

\begin{figure*}
    \centering
    \includegraphics[width=1\textwidth]{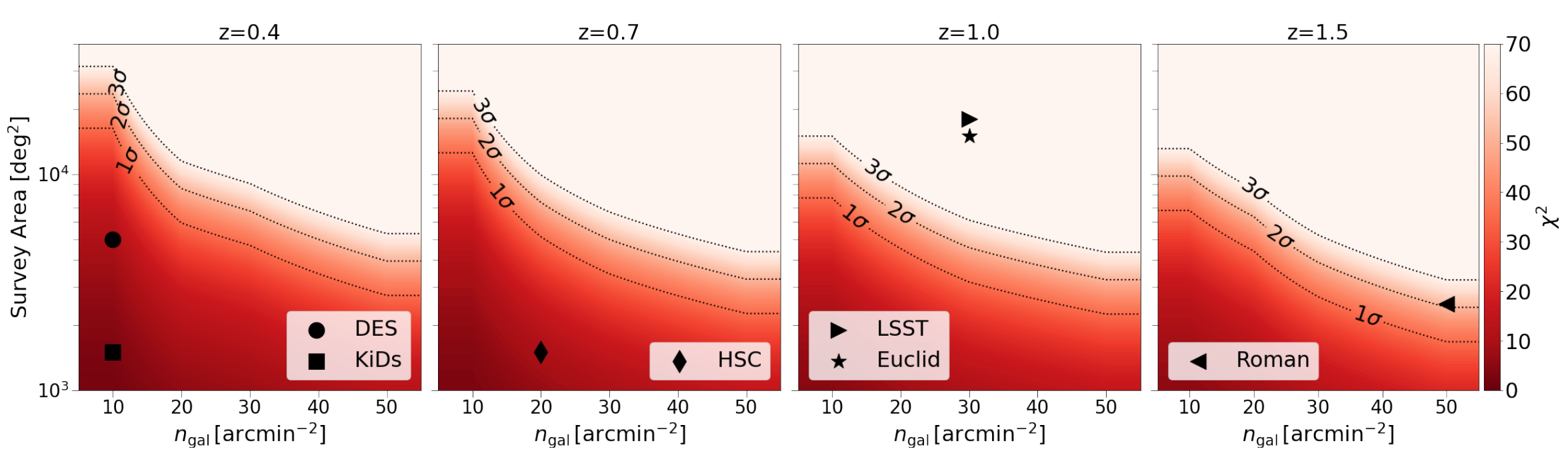}
\caption{The significance of the difference between the peak counts in the $\kappa$BCM and the full hydrodynamical $\kappa$TNG maps, as a function of shape noise ($n_{\rm ngal}$ and survey area.  The four panels roughly correspond to the median redshifts in  existing/ongoing and future WL surveys, as labeled in each panel. The symbol marking each survey also indicates the approximate depth and area of that survey. The three curves in each panel mark the 1, 2, and 3$\sigma$ uncertainties of the $\kappa$TNG model. The $\kappa$BCM predictions are within these uncertainties of most surveys, with the exception of LSST and {\it Euclid}. The small statistical errors in these two large future surveys render the $\kappa$BCM predictions discrepant well beyond the 3$\sigma$ level.}
    \label{fig:survey}
\end{figure*}

In Fig.~\ref{fig:survey} we show a map of the resulting $\chi^2$ values as a function of depth ($n_{\rm gal}$) and survey size.  The lines mark the 1, 2, and 3$\sigma$ contours, and the four panels correspond to increasing redshifts, with the relevant surveys marked in each panel areas of KiDS, DES, HSC, {\it Euclid}, and {\it Roman}.  This plot shows that the BCM is sufficient to use for most surveys considered in this work. In particular, at the depth/area combinations of the existing/ongoing surveys DES, KiDs and HSC, the BCM predictions are well within the 1$\sigma$ uncertainties.  The deepest survey we consider, {\it Roman}, still remains within the 2$\sigma$ uncertainty.   On the other hand, in the two largest surveys, LSST and {\it Euclid}, the $\kappa$BCM peak counts become discrepant from $\kappa$TNG well beyond the $3\sigma$ level.  This is clearly due to the large number of peaks in these surveys, which correspondingly strongly reduces the statistical errors on the peak counts. In the next section, we attempt to alleviate these discrepancies between $\kappa$TNG and $\kappa$BCM by removing bins from consideration. 

\subsection{Differences between $\kappa$BCM and $\kappa$TNG peak counts} \label{sec:peak_locs_and_heights}

We next take a closer look at how the population of peaks in $\kappa$BCM and $\kappa$TNG differ, on the map level.

Reexamining Fig.~\ref{fig:kappa_diffs}, we find regions in the leftmost panel containing multiple high density peaks (for example, the region bounded by $0.5\leq X\leq 1.5\,\mathrm{deg}$ and $3\leq Y\leq 4\,\mathrm{deg}$). These peaks appear to coincide with alternating \emph{blue-red} ``dipole'' patterns in the right two panels. A natural explanation for this pattern is that baryonic effects shift the location of the peaks on the $\kappa$TNG maps, compared to  $\kappa$Dark. The locations of haloes in dark-matter only {\it vs.} hydrodynamical simulations are generally different. Since the BCMs change only the halo profiles, but not their locations, they are unable to capture this baryonic effect.  

\begin{figure}
    \centering
    \includegraphics[width=0.5\textwidth]{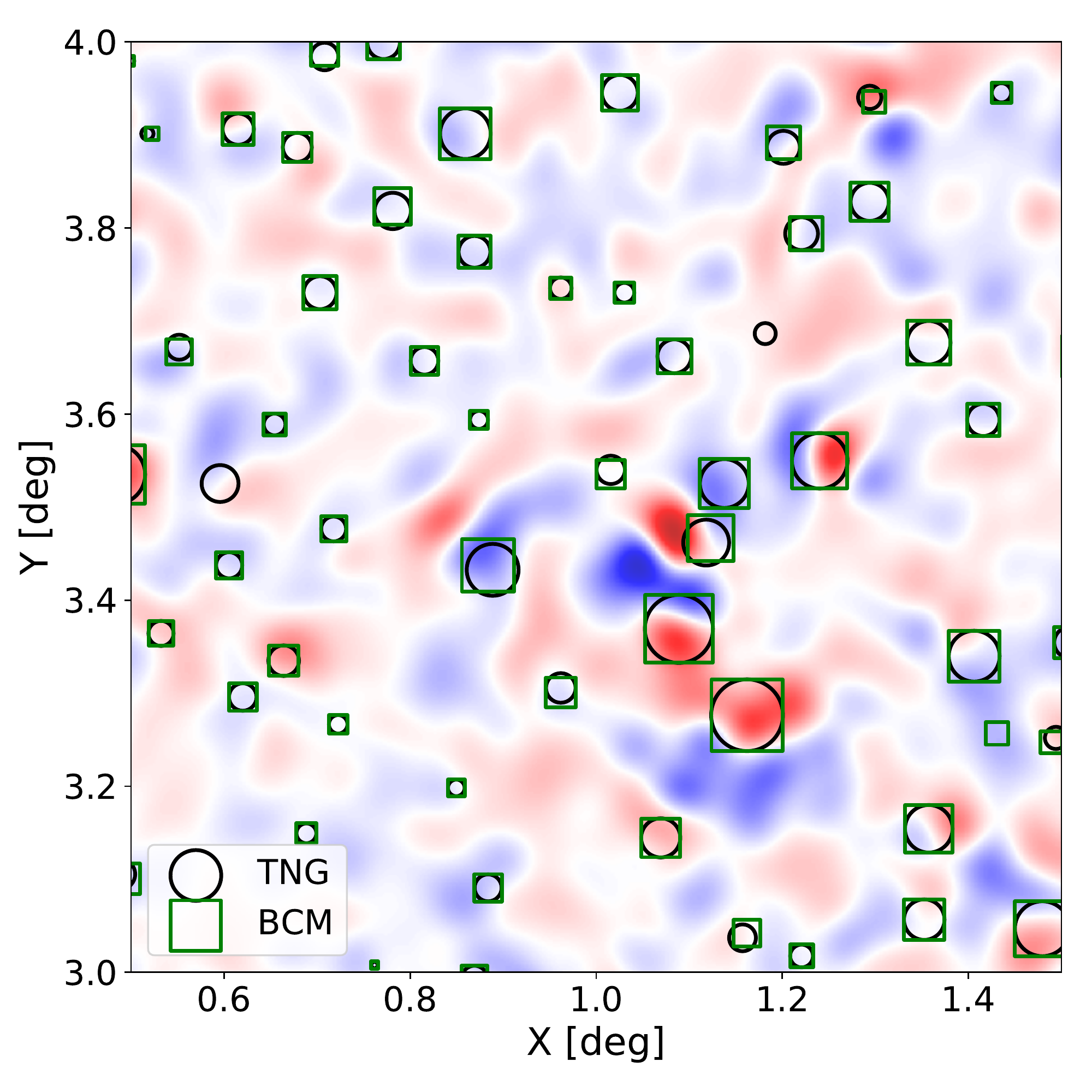}
    \caption{A subset of peaks, in a zoom-in of a small region, of the difference $\kappa$ map ($\kappa$TNG-$\kappa$BCM, where red is positive and blue is negative) in the rightmost panel of Fig.~\ref{fig:kappa_diffs}. The size of each circle (square) is proportional to the height of the peak in $\kappa$TNG ($\kappa$BCM).  The figure illustrates the small offset between the peak centres in $\kappa$TNG and $\kappa$BCM evident in $>25\%$ of the peaks, causing an alternating colour ``dipole'' pattern. We interpret this as being caused by different halo centre locations in the TNG and BCM density fields (see $X \approx 1.15$, $Y\approx 3.5$).  The BCM cannot reproduce baryonic effects well outside haloes, or the impact of baryonic effects on the halo locations. }
    \label{fig:peak_loc_diff}
\end{figure}
\begin{figure}
    \centering
    \includegraphics[width=0.5\textwidth]{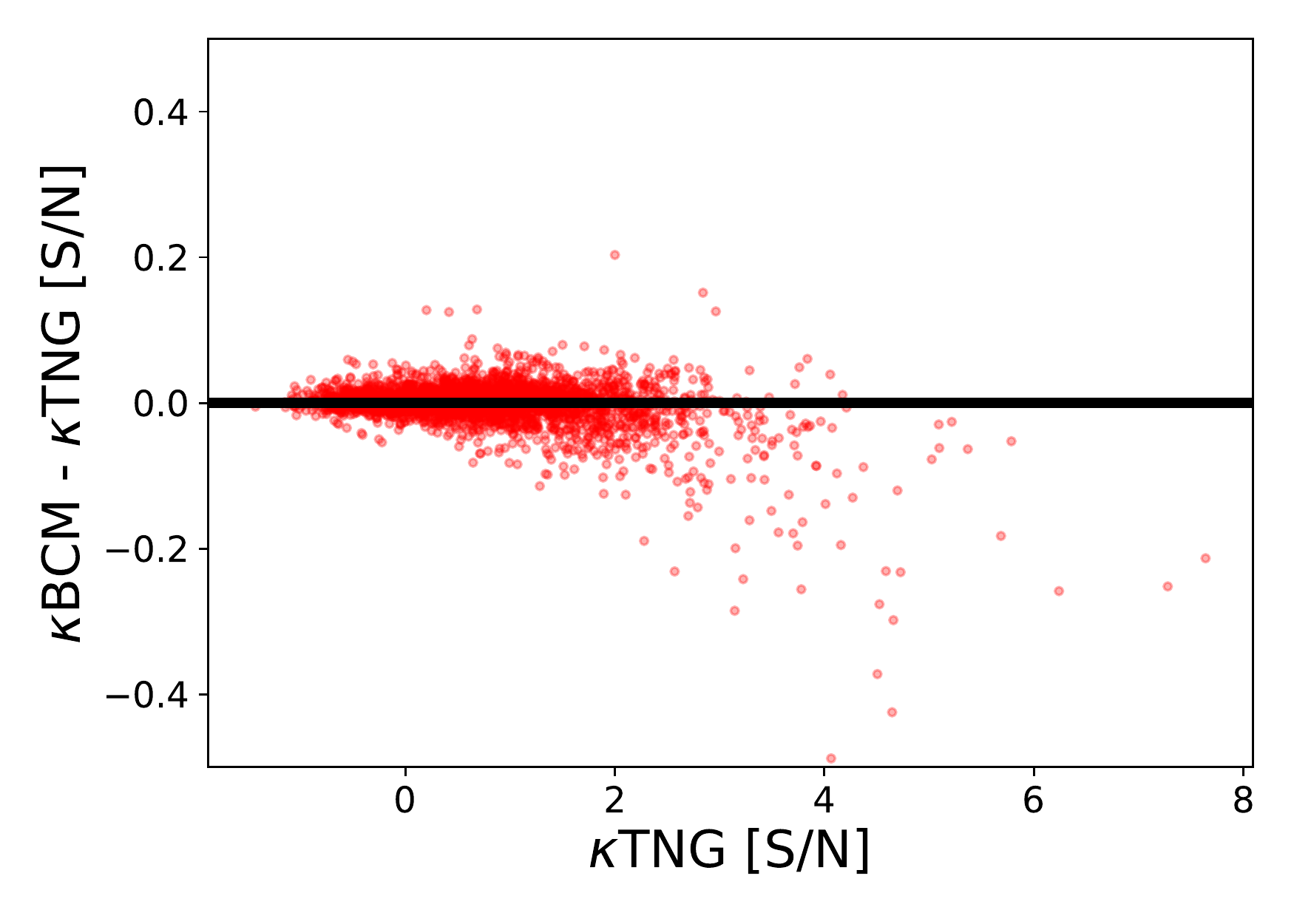}
    \caption{The difference between peak heights for $5$ $\kappa$BCM and $\kappa$TNG maps (corresponding to $3337$ peaks) at $z=1.5$ plotted against the peak heights from $\kappa$TNG. We see a systematic under-prediction by $\kappa$BCM of peak heights at ${\rm S/N}\gtrsim 4$. }
    \label{fig:peak_height_diff}
\end{figure}
To test this hypothesis further, in Fig.~\ref{fig:peak_loc_diff} we show the population of peaks in a small sub-region of the convergence maps from $\kappa$TNG and $\kappa$BCM. The size of each circle (square) is proportional to the $\kappa$ value of the peak in $\kappa$TNG ($\kappa$BCM). The background colours reproduce the rightmost panel of Fig.~\ref{fig:kappa_diffs}, representing $\kappa$TNG-$\kappa$BCM. Fig.~\ref{fig:peak_loc_diff} shows that most of the peaks exist in both $\kappa$TNG and $\kappa$BCM, and have similar $\kappa$ values. On the other hand, the heights do not match perfectly, and there are two $\kappa$TNG peaks that are missed in $\kappa$BCM. Furthermore, the centres of many peaks are visibly offset.  A clear example of this is the peak at $X\approx1.3\,\mathrm{deg}$, $Y\approx3.55\,\mathrm{deg}$. The square ($\kappa$BCM peak) is shifted to the left, while the circle ($\kappa$TNG peak) is shifted to the right. This  matches the dipole colour pattern of the background, which shows a $\kappa$BCM-dominated region to the left of a $\kappa$TNG-dominated region. When comparing the locations of peaks in general, we find that $\gtrsim 25\%$ of the peaks in $\kappa$TNG and $\kappa$BCM have offsets by at least one pixel.

A plausible origin of this effect is that the BCM is unable to account for the formation histories and locations of dark matter haloes. While baryonic effects, such as AGN feedback, are implemented at a single snapshot in space and time by the BCM, these effects are not propagated through the simulation, resulting in different halo centres. Additionally, baryonic processes can modify the density distribution well outside haloes, as well, beyond the regions captured by the BCMs.  Since non-halo matter is known to contribute to peak counts~\citep{sabyr21}, this effect can also contribute to changing peak locations and heights, when compared to dark-matter only simulations. 

Beyond peak placement, the BCM assigns systematically lower $\kappa$ values to peaks. As shown in the right panels of Fig.~\ref{fig:peak_hist} and Fig.~\ref{fig:BCM_TNG}, the number of peaks in the highest S/N regions are underestimated by the BCM when compared to either $\kappa$Dark or $\kappa$TNG maps.  Further evidence for this can be seen in Fig.~\ref{fig:peak_loc_diff}. Regions with high peaks that share the same peak centre (e.g. $X \approx 1.18\,\mathrm{deg}$, $Y=3.3\,\mathrm{deg}$) tend to be redder, corresponding to a larger $\kappa$TNG peak height. This is likely due to the BCM over-accounting for AGN feedback, an effect which we commented on in \S~\ref{sec:Results}. This effect would disperse more matter out of haloes, reducing their density and lower the peak $\kappa$ values.  As the peak-count histograms show, their slope becomes steeper at higher S/N.  Because of this, we expect that the number of these high peaks (${\rm S/N}\gtrsim 4$), are the most affected by moving peaks into lower-$\kappa$ bins. 

This motivates us to investigate the peak heights more carefully. We find all the peaks that are present in both $\kappa$TNG and $\kappa$BCM for 10 individual maps, where we identify the matching between the peaks in both maps by their distance being smaller than a threshold of $2.9\,{\rm arcmin}$. In Fig.~\ref{fig:peak_height_diff} we compare peak heights from $\kappa$TNG and $\kappa$BCM. 
From this representation it is clear that the BCM systematically assigns lower peak heights on average
at ${\rm S/N}\gtrsim 4$. Intuitively, the increased AGN feedback present in BCM will manifest as smaller peak heights for the peaks that are made up of individual massive haloes and which have expelled an overestimated amount of matter. For smaller peaks which rely on multiple low-mass haloes and are therefore less sensitive to to this effect, we see that the systematic underestimation in the peak heights of the BCM are less severe or non-existent.

We also note that there is a small number of peaks in the $\kappa$TNG maps that do not have a clear match in $\kappa$BCM and vice versa. More specifically, the number of unmatched peaks that we had to exclude in Fig.~\ref{fig:kappa_diffs} is roughly 2\% of the total $\kappa$TNG peaks. Likewise, there are peaks in $\kappa$BCM without a match in $\kappa$TNG, which we find also account for about 2\% of the total.  The asymmetry in number among this population of unmatched peaks is small, and, furthermore, the missed peaks have a maximum S/N of 3.5. We therefore conclude that the reason for BCM's underprediction of the number of high S/N peaks is the systematic underestimation of the peak heights, rather than missing peaks entirely.

\subsection{Mitigation of high S/N peak discrepancies}
\label{sec:mitigation}

We have seen that the BCM is unable to reproduce  hydrodynamical simulations at the pixel level, and that peak counts between $\kappa$TNG and $\kappa$BCM deviate at ${\rm S/N}\gtrsim 4$. Further we find that future surveys with large areas such as LSST and {\it Euclid} produce $\chi^2$ differences well above the $1\sigma$ random coincidence level. 

\begin{figure}
    \centering
    \includegraphics[width=0.49\textwidth]{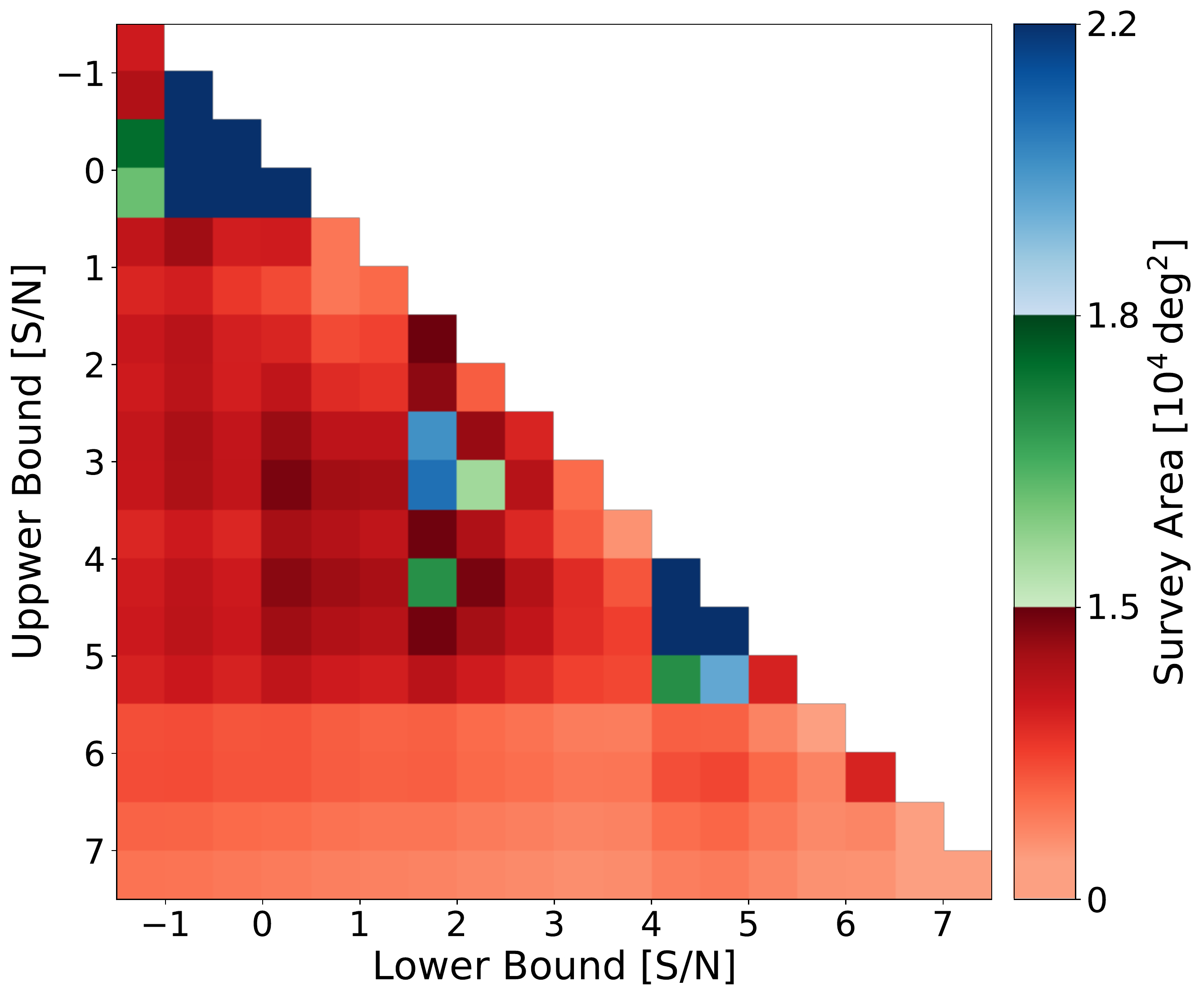}
    \caption{We compute the maximum survey area that would produce $1\sigma$ deviations between $\kappa$TNG and $\kappa$BCM, when the lowest and/or highest S/N peaks are excluded from the $\chi^2$ calculation. Each pixel represents a different combination of these cuts given by the x-axis lower bound and y-axis upper bound. We highlight the regions above {\it Euclid}'s area (\emph{green}), LSST's area (\emph{blue}) and all smaller areas (\emph{red}).
    }
    \label{fig:SN_cut}
\end{figure}

With this motivation, we consider a remedy to lower the value of $\chi^2$ for LSST and {\it Euclid} from $\chi^2_\mathrm{LSST} = 77.98$ and $\chi^2_{Euclid}=64.97$ to the $1\sigma$ threshold $\chi^{2*}_{\rm D.O.F}$ which depends on the number of bins. To do this, we remove the most discrepant peaks and use only the remaining subset of histogram bins.  Since both the highest and lowest peaks are highly discrepant, we consider various combinations of lower-S/N and upper-S/N cuts in the $\chi^2$ computation.  In Fig.~\ref{fig:SN_cut}, we show the maximum allowable survey area corresponding to $1\sigma$ $\chi^{2*}_{\rm D.O.F}$ as a function of these S/N thresholds. Each pixel corresponds to a S/N cut, with the bottom left representing utilizing the full $18$ bins, and the diagonal elements containing only a single bin. \emph{Green} and \emph{blue} pixels show cuts that generate a $1\sigma$ maximum allowable survey area greater than or equal to {\it Euclid} and LSST respectively. As the figure shows, additional bins can either decrease or increase the allowed area. This is because bins in which BCM is accurate decrease $\chi^2$ with respect to $\chi^{2*}_{\rm D.O.F}$ (accounting for the extra D.O.F) , while discrepant bins cause an increase. This figure shows that reducing $\chi^2$ to the $1\sigma$ level for LSST and {\it Euclid} is not possible by just excluding the highest and/or lowest peaks.  The $1\sigma$ criterion can be met only by removing the vast majority of peaks from the analysis, which would mean a significant loss of cosmological sensitivity. This result could also be anticipated from the bottom panel in the $n_{\rm gal} = 30$, $z=1.0$ case in Fig.~\ref{fig:BCM_TNG}. This shows that the discrepancy between $\kappa$BCM and $\kappa$TNG is largest at ${\rm S/N}\gtrsim 4$, but the lower-S/N bins still contribute significantly. 

As we saw in Figs.~\ref{fig:kappa_power}~and~\ref{fig:peak_height_diff}, there is a systematic under-prediction of peak heights and power spectra. We speculated that this was due to an overestimation of the AGN feedback component, thus removing too much matter from the most massive haloes.  This suggests a different mitigation approach: to change the BCM model parameters or possibly modify the BCM model itself. \cite{Arico+2021} shows that fitting the BCM parameters jointly to the power spectrum and the bispectrum provides a vast improvement to the small-scale bispectrum in hydrodynamical simulations, compared to using BCMs fit only to the power spectrum. Furthermore, they find that this joint fitting does not significantly degrade the fit to the power spectrum. In future work, we plan to fit the BCM parameters to the peaks themselves, which we expect would increase the agreement between peak heights at the largest peaks. Another possible alternative is to change the BCM model to provide a more flexible AGN feedback component that is reduced for the more massive halos, removing the systematic under-prediction for the largest peaks.

\section{Summary and conclusions}
\label{sec:conclude}

In this work, we investigated the level of agreement in weak lensing peak counts between semi-analytical baryonic correction models (BCMs), and hydrodynamic simulations. We applied a BCM developed in \cite{Arico2020}, which has been shown to match 3D matter power spectra and lensing convergence power spectra in IllustrisTNG, and other hydrodynamic simulations. After verifying that our implementation and parameterisation is comparable to the results of \cite{Arico2020}, we performed ray-tracing as developed in \cite{Osato21}, and computed the peak statistics from the resulting convergence ($\kappa$) maps. 

We find that the ray-traced BCM maps ($\kappa$BCM) is able to reproduce the peak statistics from ray-traced TNG maps ($\kappa$TNG) at $\sim$ percent accuracy for a range of redshifts, survey depths, and sky areas up to a peak height of ${\rm S/N}\approx 4$, but significantly under-predicts the number of peaks above this threshold. We investigated the pixel-level agreement between $\kappa$BCM and $\kappa$TNG maps, and found that while $\kappa$BCM matches almost all of the $\kappa$TNG peaks, it yields systematically lower peak heights, misses a small fraction of the peaks, and yields a small offset in the peak locations. We believe that the latter effect is due to TNG capturing the impact of baryons well outside haloes and halo formation histories affected by baryons that the BCM cannot account for.  The most significant effect for the peak-count statistic arises from $\kappa$BCM's underestimation of the heights of the largest-S/N peaks. We speculate that this is caused by the BCMs over-accounting for AGN feedback and removal of excess mass from haloes, and suggest that this could be corrected by further modifying the BCMs.

We investigated the statistical significance between the $\kappa$BCM and $\kappa$TNG peak histograms, using a simple $\chi^2$ statistic. We found that deviations between $\kappa$BCM and $\kappa$TNG are statistically insignificant at the depths and sky coverages typical of current and on-going surveys (well within $1\sigma$ uncertainties).  On the other hand, the discrepancy rises to the $\approx 2\sigma$ level for a deep survey such as {\it Roman}, and becomes even more significant (well above $3\sigma$) for large survey areas, such as in the surveys by LSST and {\it Euclid}. We considered a simple remedy in which we excluded the most discrepant lowest and/or highest peaks from the $\chi^2$ computation, but found that this was insufficient in reducing $\chi^2$ to the level of $1\sigma$.

In this work, we did not perform any optimization by parameter fitting for the BCM using the peaks, and instead adopted the values presented by \cite{Arico2020} that best fit the 3D matter power spectrum. We will investigate the differences in parameters when fitting power spectra and peaks in a future work, to assess whether a single set of parameters simultaneously fits both statistics, or if peak-count predictions could be further improved through calibration of BCM parameters. The influence of BCMs on other non-Gaussian statistics, such as WL minima and Minkowski functionals, as well as using more realistic survey parameters (such as redshift distributions) are also of interest, and will be also explored.

Most importantly, in this work, we studied only the statistical difference between peaks counts in $\kappa$BCM and $\kappa$TNG. In future work, we will assess the impact of BCM biases on cosmological parameter estimation.  Our results presented here are somewhat conservative, in the sense that when the peak counts are statistically indistinguishable, they can not bias cosmological parameter-inference.   On the other hand, when the $\kappa$BCM peak counts statistically differ from $\kappa$TNG, they may or may not cause a significant cosmological-parameter bias, depending on how the cosmological {\it vs.} BCM parameters impact the peak counts.

\section*{Acknowledgements} 
We acknowledge support by NASA ATP grant 80NSSC18K1093 (to ZH), the use of the NSF XSEDE facility Stampede2, and the Columbia University High-Performance Computing cluster Ginsburg for simulations and data analysis used in this study.
KO is supported by JSPS Research Fellowships for Young Scientists.
This work was supported by Grant-in-Aid for JSPS Fellows Grant Number JP21J00011.

%%%%%%%%%%%%%%%%%%%%%%%%%%%%%%%%%%%%%%%%%%%%%%%%%%
\section*{Data Availability}
The data underlying this article were accessed from Stampede2. The derived data generated in this research will be shared on reasonable request to the corresponding author.

%%%%%%%%%%%%%%%%%%%% REFERENCES %%%%%%%%%%%%%%%%%%

% The best way to enter references is to use BibTeX:

\bibliographystyle{mnras}
\bibliography{biblio} % if your bibtex file is called example.bib

%%%%%%%%%%%%%%%%%%%%%%%%%%%%%%%%%%%%%%%%%%%%%%%%%%

%%%%%%%%%%%%%%%%% APPENDICES %%%%%%%%%%%%%%%%%%%%%

% \appendix

% \section{Some extra material}

% If you want to present additional material which would interrupt the flow of the main paper,
% it can be placed in an Appendix which appears after the list of references.

% %%%%%%%%%%%%%%%%%%%%%%%%%%%%%%%%%%%%%%%%%%%%%%%%%%

% Don't change these lines
\bsp	% typesetting comment
\label{lastpage}
\end{document}